# Two-Dimensional Quantum Search Algorithm

Arti Chamoli and Samina Masood

Department of Physics, University of Houston-Clear Lake,
Houston, Texas-77058

**Abstract**

Quantum mechanical search induces polynomial speed up in an unsorted database search process. In case of classical linear search the computational time increases with the dimensionality of the query. However, quantum parallelism, inherent to quantum systems, does not let multi-dimensional query processing affect the computational time of the quantum search algorithm. In this letter, a two-dimensional search process has been proposed. It has been shown that a two-dimensional search process can be accomplished without increasing the computational time due to implicit quantum parallelism.

PACS: 03.67.Ac, 03.67.Lx, 03.65.Ud, 03.67.-a

Quantum mechanics has added a new dimension to the world of computation. The merger of quantum mechanics into classical computation has evolved the unique field of quantum computation. In principle, quantum computer can not only out perform a classical computer in terms of computational time but can perform several tasks that are inconceivable for a classical computer. This is owing to exploitation of unique quantum features like massive parallelism and entanglement in the processing of quantum computers. The algorithms based on the laws of quantum mechanics have further strengthened the concept of quantum computation. Grover's quantum search algorithm [1] is one such algorithm that exemplifies the uniqueness of quantum computation. It is a search algorithm for an unsorted database that provides quadratic speed up in the search process over classical linear search. The searching is accomplished by enhancing the probability amplitude of the queried term. Amplitude amplification herein manifests the probabilistic nature of quantum mechanics.

The quantum search algorithm has been studied in context of various theoretical and experimental aspects, ever since its inception. The experimental implementation of the algorithm using Nuclear Magnetic Resonance (NMR) mechanism [2], optical systems [3] or trapped ions method [4] etc. has been extensively studied by the researchers. The search algorithm has also been studied with respect to adiabatic evolution of quantum systems [5] and non-adiabatic as well [6]. Issues like quantum decoherence [7] and quantum circuitry errors [8] which come across the feasibility of the search algorithm have also been addressed by the researchers. In addition, the algorithm has also been studied as continuous time evolution of a quantum system [9]. The Grover's search algorithm has been studied as an operational measure for quantification of entanglement in quantum states [10]. It has also been examined in context of classical [11] and quantum cryptography [12].

The original search algorithm considers searching for a single state as the query term. However the complete search process is likely to have more than one query terms. Searching for more than one query term is a unique feature of quantum search algorithm because in classical searching only single query can be processed at a time. So far multi object search by Grover's search algorithm has been accomplished by considering more than one marked state as query [13] in which case the size of the search operator, specifically of the phase flip operator, increases. This in turn increases the query complexity. In this paper, we aim to design a multi object quantum search algorithm. The original Grover's algorithm has been modified for this study.

We start with a brief review of original Grover's search algorithm. The algorithm starts with state initialization in which an n-qubit register is prepared in the state $|0\rangle^{\otimes n}$ and a Hadamard gate $H = \frac{1}{\sqrt{2}}\begin{pmatrix} 1 & 1 \\ 1 & -1 \end{pmatrix}$ is applied to each qubit. The application of Hadamard gate transforms the state of the register to a uniform superposition state of the form

$$|\Psi\rangle = \frac{1}{\sqrt{N}} \sum_{a=0}^{N} |a\rangle$$

The state $|\Psi\rangle$ is then considered as the initial state for the algorithm to proceed. The base states of the superposition state $|\Psi\rangle$ represent the items of the database. Thus an n-qubit quantum register will have $N=2^n$ elements in its database. The second step is to invert the phase of the state marked as the query term. This is achieved by applying the quantum phase flip operator, $I - 2|\tau\rangle\langle\tau|$, on the uniform superposition state $|\Psi\rangle$. $|\tau\rangle$ is the marked state. This is followed by another quantum operator, inversion about mean of the amplitudes of all the base states. The operator is of the form $2|\Psi\rangle\langle\Psi| - I$. These two operators are applied consecutively for a finite number of times before the measurement to obtain the marked state is made so as to maximize the accuracy of measurement.

In this modified Grover's search algorithm, the search process starts using two different atoms $A_1$ and $A_2$ (say). The two atoms $A_1$ and $A_2$ are encoded as qubit 1 and qubit 2. The encoding is done in the energy states of the atom and the nuclear spin states as well. Since the atom can be excited from ground state to a higher energy state, it can serve the purpose of a qubit. Similarly, the two possible spin orientations up and down, of the nucleus qualify it as a qubit. Quantum mechanically, the energy states and the nuclear spin states are mutually commuting degrees of freedom so they can be measured simultaneously without affecting each other. Therefore total wave function can be considered as a superposition of all possibilities for both. Thus it is possible to encode two different entities on the energy and spin states.

The search process in this case is explained with the help of an example. We have access to a database of census of an area which has the record of name, educational qualification and gender of each person living in that area. The database is organized with names in alphabetical order. Thus if we have to fetch the names of the females with master's degree living in that area, it becomes an unsorted database search. Classically, the search process would start by querying through the database for the female gender and then from among the entries for female genders, the computer would search for the master's degree; alternatively, the computer will first search for the entries with master's degree and then from among the master's degree holders it will search for the female gender. The computational time can be managed to some extent by the order of processing of query terms, only if we have an idea of distribution of both query terms in the database. This being a linear search will have the time complexity proportional to the size of the database. The schematic representation of linear classical search in above mentioned two possible ways is shown in Figure 1(a) and Figure 1(b).

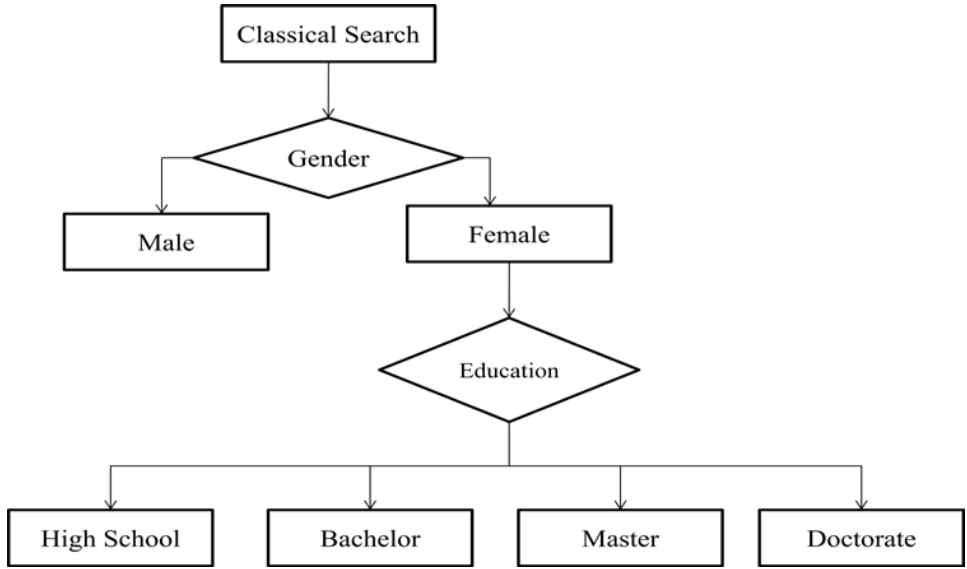

**Figure 1(a) Classical linear search with gender as the first query.**

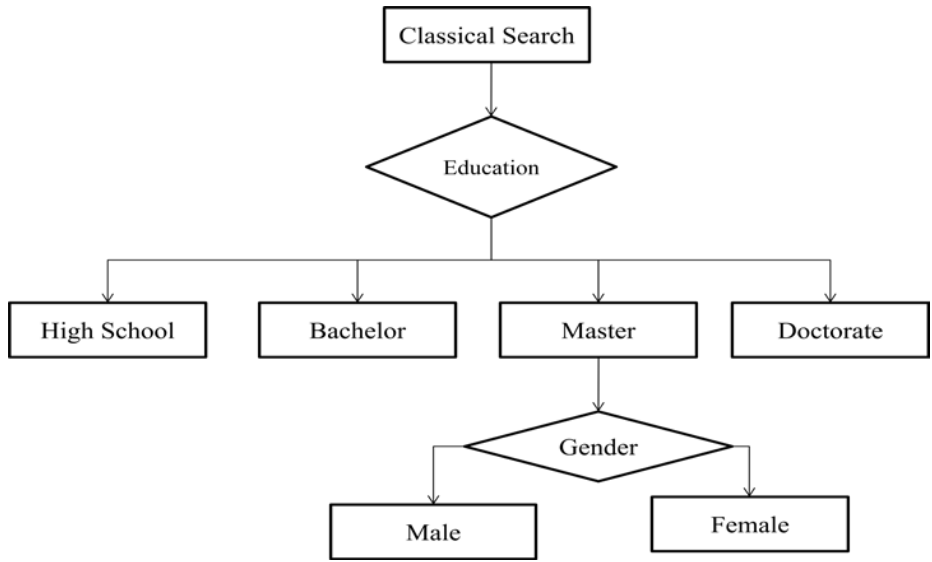

**Figure 1(b) Classical linear search with education as the first query.**

In case of quantum search, the two atoms $A_1$ and $A_2$ are considered in ground energy state as $|G_i\rangle$ (for i = 1, 2 for atom $A_1$ and $A_2$ respectively). Ground state of atom $A_1$ is encoded as female, $|G_1\rangle$, and that of atom $A_2$ as male, $|G_2\rangle$. Unlike energy states, the four possible nuclear spin states of the two atoms in superposition are exploited to encode four educational degrees. It is assumed that the atom $A_1$ ($A_2$) has $|0_1\rangle$ ($|0_2\rangle$) as the spin-up state and $|1_1\rangle$ ($|1_2\rangle$) as the spin down state. Thus the four possible combinations can be encoded as $|0_1 0_2\rangle \Leftrightarrow$ High school, $|0_1 1_2\rangle \Leftrightarrow$ Bachelor, $|1_1 0_2\rangle \Leftrightarrow$ Master and $|1_1 1_2\rangle \Leftrightarrow$ Doctorate.

Energy and nuclear spin being commuting degrees of freedom, the state of the composite system can thus be written as a superposition of all possible states for both:

$$|\Psi\rangle = \sum_a \left( \psi_1(a)|G_1, a\rangle + \psi_2(a)|G_2, a\rangle \right)$$

where $|a\rangle$ represents the superposed nuclear spin states of the atoms $A_1$ and $A_2$

$$|a\rangle = \frac{1}{2}\left( |0_1 0_2\rangle + |0_1 1_2\rangle + |1_1 0_2\rangle + |1_1 1_2\rangle \right)$$

Now for the search process, we will perform two different operations on the composite system simultaneously. We will consider the two modes of a photon, mode A and mode B (say), with energy equivalent to the excitation energy of the two atoms. The two photon modes are such that the mode A is absorbed exclusively by atom $A_1$ and mode B by atom $A_2$. Any other interaction is not sustained by the composite system. For the query purpose, photon in mode A is encoded as female and that in mode B as male. Thus photon in mode A will excite the atom $A_1$ from state $|G_1\rangle$ to state $|E_1\rangle$ and the one in mode B will excite the atom $A_2$ from state $|G_2\rangle$ to state $|E_2\rangle$. Figure 2(a) and Figure 2(b) show the change in energy states of atoms $A_1$ and $A_2$ with the absorption of photon modes A and B respectively.

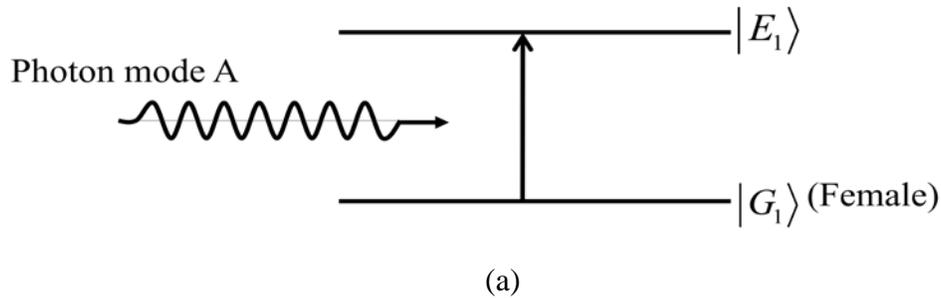

(a)

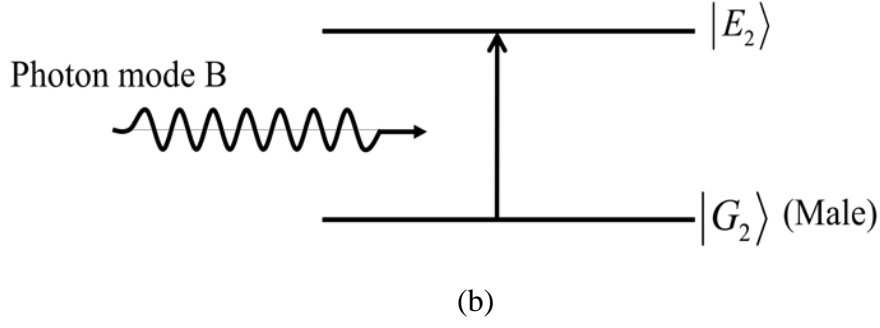

(b)

**Figure 2** Atom $A_1$ ($A_2$) jump to state $|E_1\rangle$ ($|E_2\rangle$) from state $|G_1\rangle$ ($|G_2\rangle$) with the absorption of photon mode A (B).

Going by our example, in order to retrieve the names of the females with master's degree, photon in mode A is made to interact with the atomic system. In the process, it is absorbed by atom $A_1$ and the atom from state $|G_1\rangle$ jumps to state $|E_1\rangle$. The state of the composite system now becomes

$$|\Psi\rangle' = \sum_a \left( \psi_1(a)|E_1,a\rangle + \psi_2(a)|G_2,a\rangle \right)$$

Thus the atomic energy state $|G_1\rangle$ encoded as female will jump to an excited energy state $|E_1\rangle$.

The query for the educational degree is made using conventional 2-qubit Grover's search algorithm on the superposed nuclear spin states of the atoms $A_1$ and $A_2$.

Finally making a measurement on the atom in the excited state $|E_1\rangle$ and a joint measurement on the nuclear spin states of the two atoms $A_1$ and $A_2$ will retrieve the name of all the females with master's degree.

This is worth-mentioning that the proposed quantum search algorithm with the two-dimensional search will reduce the computational time to $O(\sqrt{N})$ as compared to $O(N)$ in case of classical linear search, where $N$ refers to the number of entries in the database. In this way a larger database can be searched more efficiently due to quantum mechanical coupling between atoms and photons. Therefore, the quantum mechanical search not only tremendously reduces the computational time but also has a high precision with the increase in size of the databases. In case of quantum search, simultaneous processing of more than one dimensions of a multi-dimensional query is possible due to the commutative property of quantum operators. The search process discussed in this letter can be further generalized by including additional commuting degrees of freedom for encoding quantum bits thereby increasing the dimensionality of the query term in the search process. Quantum mechanical entanglement [14] for such systems of two atoms and two photon modes is being studied in detail and its time evolution can be used to get a better control on the search process at a later stage. The quantum search algorithm may also be generalized by a larger quantum system using generalized Jaynes-Cummings model [15] also.

## References

1. L.K. Grover, Phys. Rev. Lett. **79**, 325–328 (1997).

2. I. L. Chuang, N. Gershenfeld, and M. Kubinec, Phys.Rev. Lett. **80**, 3408 (1998); J. A. Jones, M. Mosca, and R. H. Hansen, Nature (London) **393**, 344-346 (1998).

3. P. Walther, K. J. Resch, T. Rudolph, E. Schenck, H. Weinfurter, V. Vedral, M. Aspelmeyer, and A. Zeilinger, Nature (London) **434**, 169-176 (2005); P. G. Kwiat, J. R. Mitchell, P. D. D. Schwindt, and A. G. White, J. Mod. Opt. **47**, 257 (2000).

4. K.-A. Brickman, P. C. Haljan, P. J. Lee, M. Acton, L. Deslauriers, and C. Monroe, Phys. Rev. A **72**, 050306(R) (2005); M. Feng, Phys. Rev. A **63**, 052308 (2001).

5. D. Daems and S. Guèrin, Phys. Rev. Lett. **99**, 170503 (2007); J. Roland and N. J. Cerf, Phys. Rev. A **65**, 042308 (2002).

6. A. Pèrez and A. Romanelli, Phys. Rev. A **76**, 052318 (2007).

7. H. Azuma, Phys. Rev. A **65**, 042311 (2002); Phys. Rev. A 66, 019903(E) (2002).

8. Gui Lu Long, Yan Song Li, Wei Lin Zhang, and Chang Cun Tu, Phys. Rev. A **61**, 042305 (2000).

9. E. Farhi and S. Gutmann, Phys. Rev. A **57**, 2403 (1998).

10. A. Chamoli, and C. M. Bhandari, Phys. Lett. A **346**, 17-26 (2005); Y. Shimoni, D. Shapira, and O. Biham, Phys. Rev. A **69**, 062303/1–062303/4 (2004).

11. D. J. Bernstein, in PQCrypto 2010 **36**, 73–80 (2010).

12. L. Y. Hsu, Phys Rev A **68**, 022306-022309 (2003).

13. G. Chen, S. Fulling, and J. Chen, Mathematics of Quantum Computation (R. Brylinski and G. Chen, eds.), (160, CRC Press, Boca Raton, Florida) (2002); G. Chen and S. Sun, Mathematics of Quantum Computation (R. Brylinski and G. Chen, eds.), (168, CRC Press, Boca Raton, Florida) (2002).

14. See for example, Samina Masood and Allen Miller, "A von Neumann Entropy Measure of Entanglement Transfer in a Double Jaynes-Cummings Model "[to be submitted for publication], *ibid*; arXiv: 0705.0681[quant-ph] and the references therein.

15. A. Romanelli, Phys. Rev. A **80**, 014302 (2009).